\documentclass{aa}
\usepackage{graphicx}
\usepackage{txfonts}
\usepackage{natbib}
\pdfoutput=1
\begin{document}
\title{Two-dimensional segmentation of small convective patterns in radiation hydrodynamics simulations}

\author{B. Lemmerer
\inst 1
\and D. Utz
\inst {2,1}
\and A. Hanslmeier
\inst 1
\and A. Veronig
\inst 1
\and S. Thonhofer
\inst 1
\and H.~Grimm-Strele
\inst 3
\and R. Kariyappa
\inst 4}

\institute{Institute of Physics, IGAM, University of Graz, Universit{\"a}tsplatz 5, 8010 Graz, Austria
\and Instituto de Astrof\'{i}sica de Andaluc\'{i}a (CSIC), Apdo. de Correos 3004, 18080 Granada, Spain
\and Institute of Mathematics, University of Vienna, Oskar-Morgenstern-Platz 1, 1090 Wien, Austria
\and Indian Institute of Astrophysics, Koramangala, 560034 Bangalore, India}

\date{Received: 29 March 2013/
Accepted: 6 February 2014}

\abstract{Recent results from high-resolution solar granulation observations indicate the existence of a population of small granular cells that are smaller than 600~km in diameter. These small convective cells strongly contribute to the total area of granules and are located in the intergranular lanes, where they form clusters and chains.}{We study high-resolution radiation hydrodynamics simulations of the upper convection zone and photosphere to detect small granular cells, define their spatial alignment, and analyze their physical properties.}{We developed an automated image-segmentation algorithm specifically adapted to high-resolution simulations to identify granules. The resulting segmentation masks were applied to physical quantities, such as intensity and vertical velocity profiles, provided by the simulation. A new clustering algorithm was developed to study the alignment of small granular cells.}{Small granules make a distinct contribution to the total area of granules and form clusters of chain-like alignments. The simulation profiles demonstrate a different nature for small granular cells because they exhibit on average lower intensities, lower horizontal velocities, and are located deeper inside of convective layers than regular granules. Their intensity distribution deviates from a normal distribution as known for larger granules, and follows a Weibull distribution.}{}


\maketitle

\section{Introduction}
\label{sec:Introduction}
New high-resolution solar telescopes provide us with the opportunity to study the solar photosphere in more detail than ever before. The recent achievements in high-resolution observations \citep[$\sim{0.1}$ arcsec; 1 m telescopes and beyond; NST, New Solar Telescope and GREGOR, see e.g.] []{2010ApJ...714L..31G,2010AN....331..620G,2012AN....333..796S} and the future improvements by the 4 m class ground-based high-resolution instrument ATST \citep[Advanced Technology Solar Telescope,][]{2010AN....331..609K} or the European counterpart EST \citep[European Solar Telescope,][]{2008SPIE.7012E..17C} raise interest in analyzing the solar photosphere and its convective patterns in previously unachievable detail. The granulation is a distinct feature of the solar photosphere generated by convection. To study solar granulation in detail, fully automated segmentation algorithms were developed and applied to observational data. Because the segmentation results strongly depend on the quality of the images, new telescopes with increasing resolution enable us to describe these features more precisely and might pave the way to the discovery of new convective patterns on smaller scales \citep[see e.g.][]{1986SoPh..107...11R, 1997ApJ...475..328S, 2008A&A...484L..17D}.\\

The recent discovery of granular substructures that form bright granular lanes \citep[see e.g.][]{2010ApJ...723L.180S,2011ApJ...736L..35Y} and of granular cells with scales smaller than the dominant scale of 1000 km, defined by \citet{1986SoPh..107...11R}, has again drawn attention to the topic of the solar granulation.  
\citet{2012ApJ...756L..27A} found a distinct subpopulation of smaller granular cells by studying observational data of the New Solar Telescope (NST) at Big Bear Solar Observatory (BBSO). These features, which the authors termed mini-granules, were found to be located within wide granular lanes where they form chains and clusters. How they form and dissipate is unclear. Therefore, \citet{2012ApJ...756L..27A} suggested to investigate this in more detail with numeric simulations of solar-magneto convection.\\
\begin{figure*}
	\centering
		\includegraphics[width=0.70\textwidth]{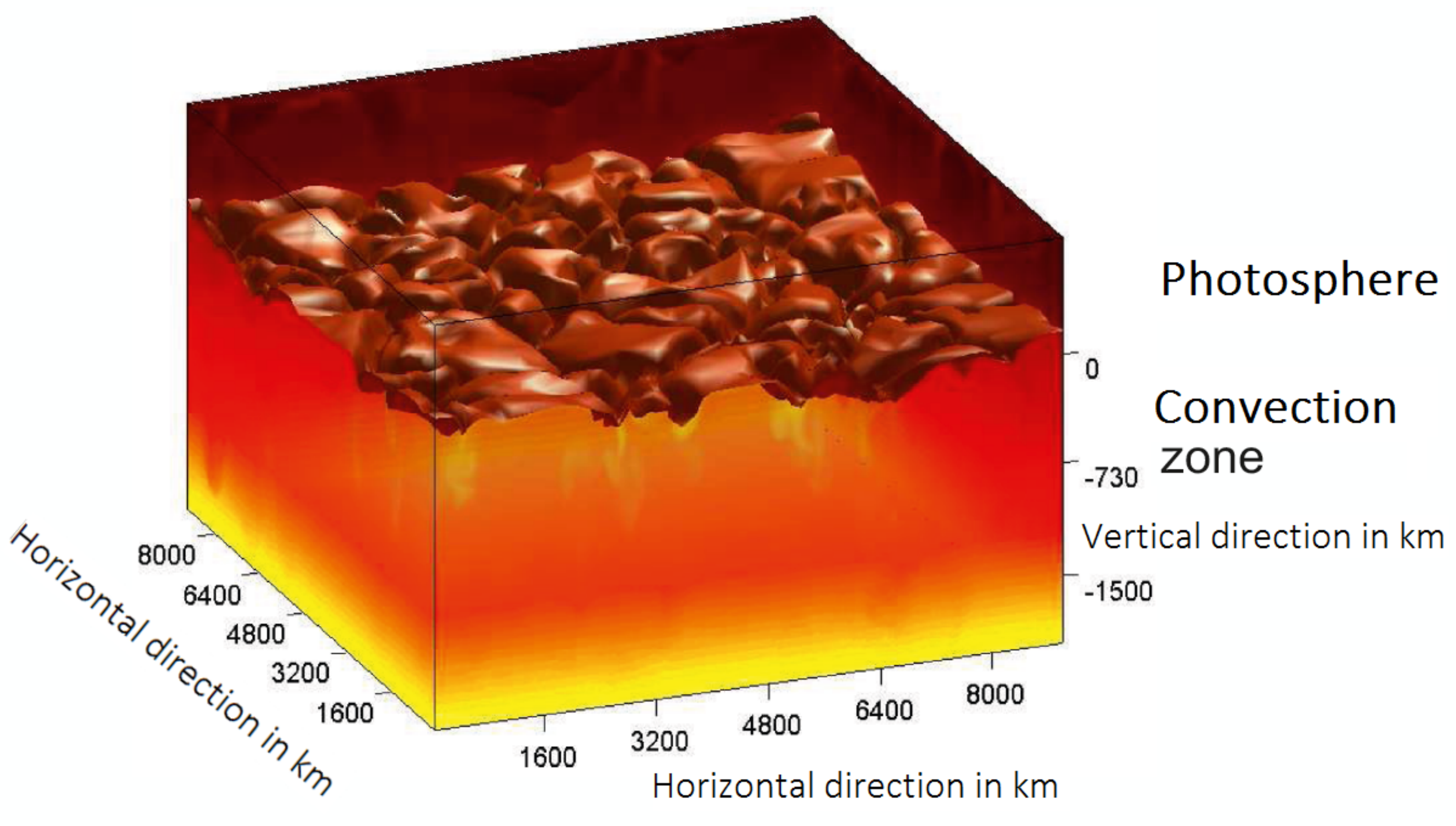}
		\caption{Snapshot of the simulated data computed by the ANTARES code during a 3D RHD  run. In this model, the temperature data are color coded. Moreover, a temperature iso-surface of 6000~K, located at a height of $\sim{750}$ km below the upper boundary, indicates the granular pattern. An artificial light source was placed above the model to accentuate the structure of iso-surface.}
	\label{cube}
\end{figure*}
 
In this study, we investigate small convective patterns with high resolution solar radiation hydrodynamics (RHD) simulations to detect and analyze convective patterns smaller than the characteristic size of the solar granulation. The advantage of RHD simulation data over magneto hydrodynamics (MHD)\footnote{More details on the state-of-the-art MHD simulations can be found in \citet{2009LRSP....6....2N}.} data regarding the analysis of the solar convection is that we do not need to consider the influence of magnetic fields on the convective pattern. For the exclusive investigation of the newly discovered mini-granules, bright grains originating from small-scale magnetic fields would be particularly interfering. Therefore, fields such as magnetic bright points \citep[MBPs, see e.g.][]{2009SSRv..144..275D,2010A&A...511A..39U} would have to be excluded from the analysis, but they inherently not present in RHD simulations.\\

For the purpose of detecting small-scale convective patterns, a new segmentation algorithm was developed. It takes several physical properties of the convection into account that are provided by the simulation, such as the vertical velocity and the emerging intensity which is defined as the outwardly directed intensity that leaves the computational box through the upper boundary. Furthermore, a clustering algorithm was developed and applied to the segmented data to study the alignment of small granules. 

\section{Simulation setup and data}
\label{sec:Data}
ANTARES (a numerical Tool for astrophysical research) is an RHD code for numerically simulating the solar near-surface convection (box-in-a-star approach) developed by \citet{2007MNRAS.380.1335M, 2010NewA...15..460M}. The  FORTRAN90-code solves the set of RHD equations using weighted essentially non-oscillatory (WENO) high-resolution numerical schemes \citep[see also][]{2013A&A...554A.119Z, 2012arXiv1209.2952M}. \\

Open boundary conditions in the vertical direction allow free in- and outflow, while in the horizontal directions periodic boundary conditions are used. A detailed description of the boundary conditions can be found in \citet{2013arXiv1305.0743G}. In the upper $\sim{30}\%$ of the simulation box the radiative heating rate is calculated using gray approximation, whereas in the rest of the box diffusion approximation is valid. The 3D model is initialized from a 1D model to which a weak perturbation is added to break the horizontal symmetries. The simulation is then thermally relaxed in an almost two-hour long period, which provides sufficient time for the development of 3D structures.\\

The box of the 3D model comprises 9 Mm in horizontal directions and 5~Mm in vertical direction. The spatial resolution in the horizontal directions is 32.1~km, in vertical direction 15.3~km, resulting in 281 by 339 grid cells. To study the granular cells we processed a data set with a temporal resolution of $\Delta t = 30$~s. Convective patterns of the simulated two hours of real solar evolution were analyzed at the calculated bottom of the photosphere.\\

Figure~\ref{cube} illustrates the temperature distribution of a 3D snapshot from an ANTARES model run. The granular pattern in the given temperature iso-surface at 6000 K is clearly visible.

\section{Automated granule detection in two dimensions}
\label{sec:Automated granule detection in two dimensions}
To analyze of the numeric RHD data we developed a two-dimensional segmentation algorithm. The automated detection of granular cells is crucial for processing large datasets and the successive statistical analysis. Several sophisticated algorithms to segment the solar granulation in observational data exist and are publicly available. Among them is the so-called multiple-level tracking (MLT) algorithm developed by \citet{2001SoPh..201...13B}, which was previously tested on ANTARES data \citep[see][]{2012CEAB...36...29L}. Our algorithm is based on the same basic idea of using multiple thresholds to segment image data, but it is also different in several crucial parts and details. The incorporation of methods of pattern recognition, such as edge-detection routines and morphological operations as well as of velocity maps of the surface flows (dopplergrams) resulted in a fast and reliable segmentation routine, which is not only useful for high-resolution data from observations, but also for those obtained from simulations. In addition, we developed a tool analyzing the clustering behavior of small granular cells.\\

To segment the granular structures in two dimensions, a reference level within the segmentation box has to be determined. To do this, we calculated the geometric height (where the optical depth $\tau = 1$) to define the bottom of the photosphere\footnote{Because we used the gray approximation, the wavelength is irrelevant.}. We evaluated the equation of optical depth for each grid point column by column from model data values. Then we extracted for each point the height at which the optical depth reaches unity. From this we were able to construct the surface of optical depth unity, which we refer to as $\tau_1$-iso-surface \citep[see also][]{2009CEAB...33...69L, 2010CEAB...34...39L}.

\subsection{Two-dimensional segmentation algorithm}
The two-dimensional segmentation algorithm was applied to the afore mentioned $\tau_1$-iso-surface. The algorithm differs in some aspects from segmentation algorithms that are based solely on intensity observations (filtergrams). Data recorded with high-resolution telescopes provide in most cases only intensity information in the form of gray-level images. Therefore, algorithms are often based on thresholding of multiple intensity levels (e.g. MLT) or on image-processing routines such as skeletonization \citep[e.g.][]{1998MmSAI..69..655F}.\\ \\
The algorithm introduced here is based on a multiple threshold-level segmentation and on image-processing techniques. The segmentation additionally profits from the use of physical quantities, that is the combination of vertical velocity and intensity or temperature that are offered by the ANTARES model. The whole scheme is a bottom-up approach. Large fragments are sequentially broken down into smaller structures by changing threshold levels, that is, changing them to higher velocities, which results in smaller structures\footnote{The opposite scheme, a top-down approach, is used in the MLT algorithm \citep[see][]{2001SoPh..201...13B}. Starting from high-intensity seeds, the granular segments grow by shifting the threshold level to lower intensity values.}.\\
\begin{figure*}
	\centering
		\includegraphics[width=1\textwidth]{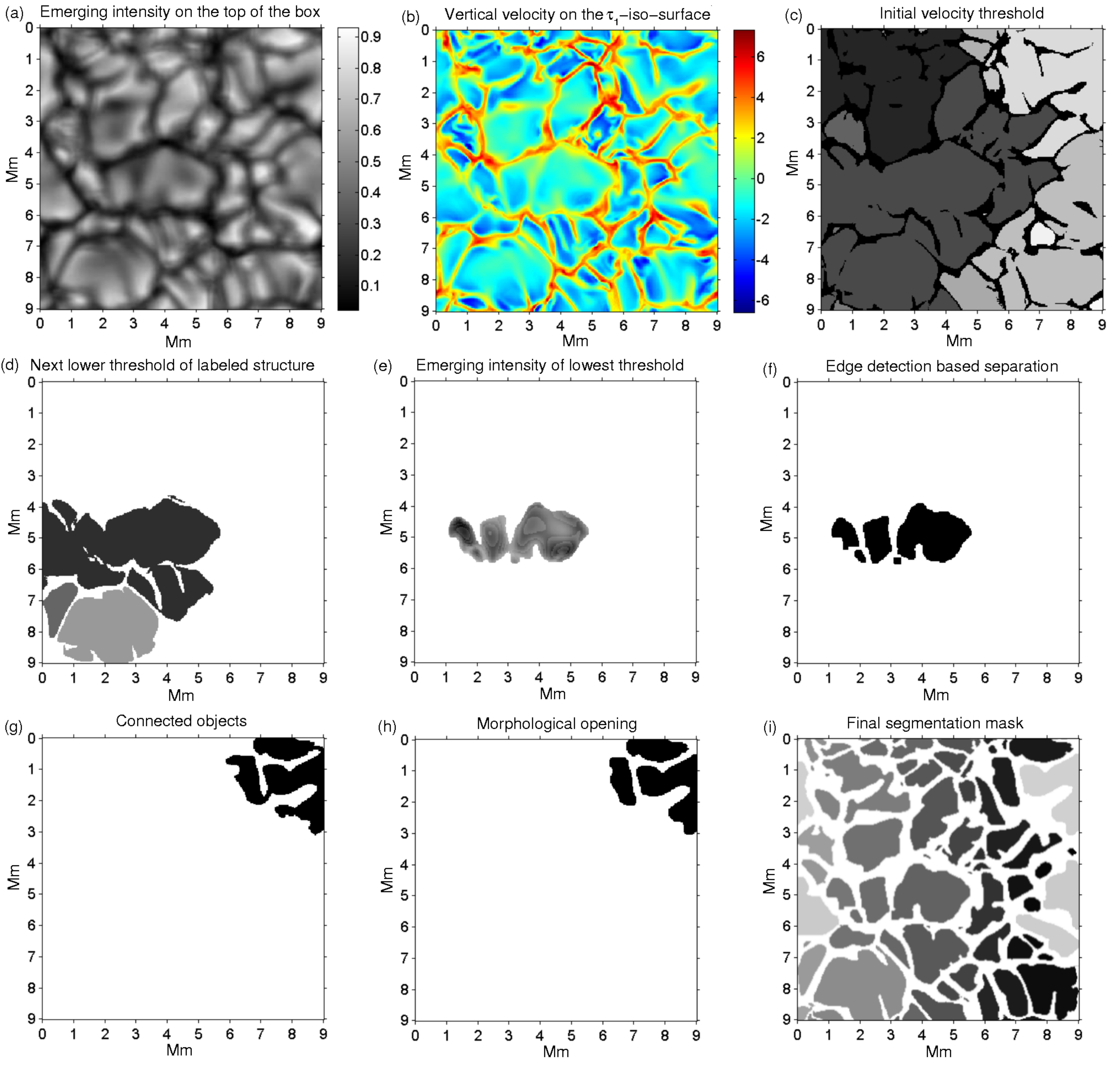}
		\caption{Steps of the applied segmentation: (a) normalized emerging intensity and (b) vertical velocity on the $\tau_1$-iso-surface of the ANTARES simulation, (c) initial vertical velocity threshold applied to the vertical velocity on the $\tau_1$-iso-surface, (d) subsequent lower threshold applied to the structure, (e) lowest threshold applied to the structure, (f) edge detection applied to the emerging intensity of the object, (g) granular cells still connected by pixels after applying of the lowest threshold, (h) morphological opening to separate connected granular cells, (i) final segmentation mask.}
	\label{Segmentation}
\end{figure*}

The algorithm starts with the pre-processing of the vertical velocity (Fig.~\ref{Segmentation}b) and the emerging intensity (Fig.~\ref{Segmentation}a). This includes normalizing both surface profiles to the interval $[0,1]$. For the velocity profile, the value 0 refers to the highest upflow velocity and 1 to the highest downflow velocity. The segmentation itself is performed by a recursive thresholding routine, which requires an initial upper threshold and a minimum lower threshold as input parameters\footnote{Because the intensity distribution varies during the whole time series, the lowest and highest segmentation thresholds are shifted by the respective change in the mean image intensity. In particular, this shifting takes care of the intensity variations caused by the 5-minute oscillation.}. While the algorithm can be applied to intensity as well as to vertical velocity profiles, for the current study we used the vertical velocity on the $\tau_1$-iso-surfaces (see Fig.~\ref{Segmentation}c) as input for the thresholding. The intensity profiles are required for later image-processing operations. The resulting structures are more distinctly separated than those obtained from the intensity segmentation. If vertical velocities are not available, intensity images can be processed instead.\\
 
After applying the initial threshold on the velocity on the $\tau_1$-iso-surface, the resulting binary image is labeled and each segmented structure is analyzed. After each thresholding step, the initial threshold is again reduced by a factor that is another input parameter. Next, this thresholding routine is applied to each of the previously found and labeled structures (see Fig.~\ref{Segmentation}d). If the previously found structure is separated because of the current thresholding, new labels are applied to the found structures. This scheme is repeated on these new structures. The thresholding routine is iterated until the lowest threshold is successfully applied (see Fig.~\ref{Segmentation}e), that is until the strongest upflows and smallest (most fragmented) structures are detected or as long as the segmented objects meet the following conditions:  
\begin{itemize}
     \item the diameter\footnote{The diameter of a granular cell is defined as the equivalent diameter of a circle with the same area as the region, computed via $\sqrt{4*A/\pi}$.} of the structure exceeds the largest diameter of granules\footnote{The largest diameter was derived from the probability density function of the granule diameters (see section \ref{sec:Results}).},
     \item The structure falls below a critical value of its solidity\footnote{The solidity of an object is defined as the ratio of its area to the area enclosed by its convex hull. A solidity of 0.7 is used as the critical value.}, indicating that granular objects are still connected.
\end{itemize}
After the final threshold was applied and objects meet the previous conditions, morphological operations are executed in the following way:
\begin{itemize}
     \item Laplacian-of-Gaussian edge-detection is applied to the profiles of the emerging intensity of the structure to detect closed contours \citep[]{2009JBO....14b9901G}, see Fig.~\ref{Segmentation}f, 
     \item a morphological opening is performed to separate objects connected by a few pixels, as illustrated in Fig.~\ref{Segmentation}g, as well as to preserve the shape of the objects \citep[]{2009JBO....14b9901G}, see Fig.~\ref{Segmentation}h.
\end{itemize}

Applying the edge-detection routine to the emerging intensity profits from the high-intensity contrast between
intergranular lanes and granular cells, which leads to a well-defined separation of merged structures. The high-intensity contrast is exploited in cases of merging of granules. When the initial phase of merging occurs in an analyzed velocity profile, the emerging intensity still shows a dark intergranular lane. These lanes indicate clear separations between granular cells. On the other hand, these granules would already appear as a merged structure in the vertical velocity on the $\tau_1$-~iso-~surface. Hence, additionally using the emerging intensity is of advantage for a correct segmentation.\\

Furthermore, incorporating the emerging intensity helps during the initial phase of the fragmentation process of exploding granules\footnote{Exploding granules are the result of the final phase of huge granules starting with a ceasing upflow in the middle of the granules. The still-expanding plasma plus the radiative cooling in the central region finally leads to a flow reversal in the middle of the granule with a downflow setting in. This downflow either creates a depression in the middle of the granule or leads to its fragmentation.}. These features often show a concavity at their center caused by downflowing plasma. In the emerging intensity these downflows appear as a depression on or outside of the granule. To prevent a loss of information regarding the statistics and physical properties of these granules, holes are automatically filled if the plasma downflow is located at the center of the granule. The structure is split into several substructures if the downflow causes a fragmentation.\\

The periodic boundaries are taken into account by excluding granular fragments, separated by the right and lower boundary (see Fig.~\ref{Segmentation}i), and appending them to the left and upper boundary, respectively. Later, this allows a calculation of the correct areas, diameters, and centroids of the granule split by the periodic model boundaries. 
The applied image-processing steps and a labeling of all objects result in the final segmentation mask (see Fig.~\ref{Segmentation}i). The border handling of split granules is illustrated by the assigned labels depicted as gray values. The mask is then applied to various profiles of physical quantities to extract statistical information of the identified granules. The final step retrieves information such as granular area, perimeter, mean intensities et cetera.

\subsection{Clustering algorithm for analyzing the alignment of small granules}
\begin{figure*}
	\centering
		\includegraphics[width=1\textwidth]{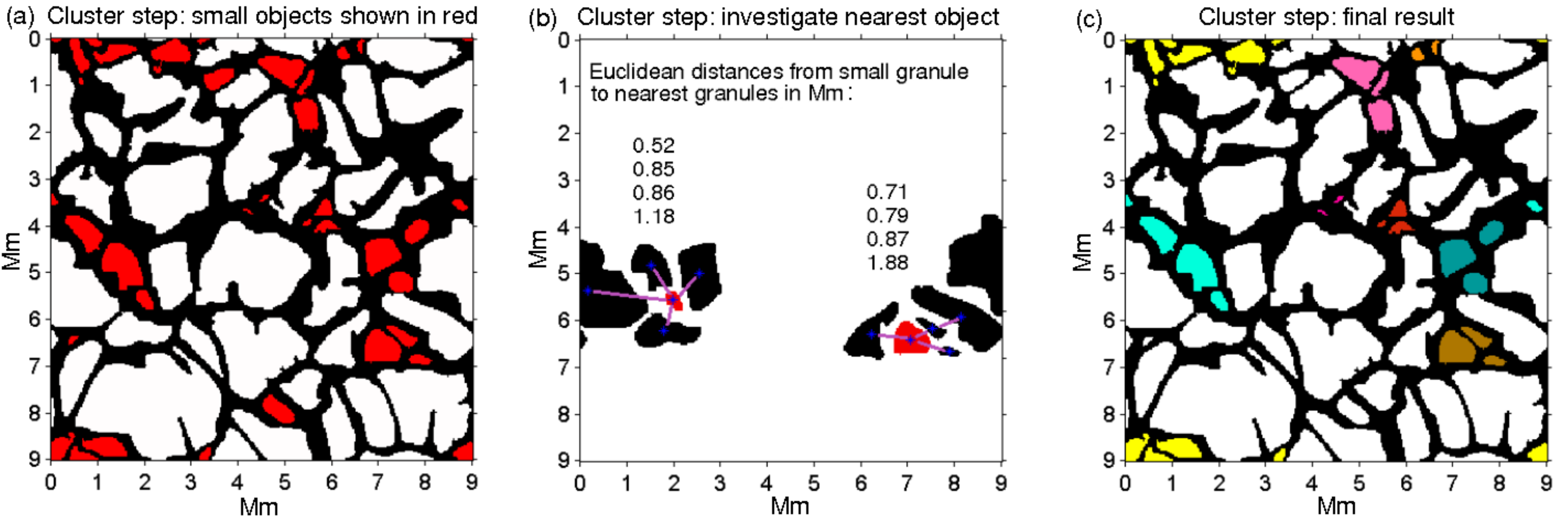}
		\caption{Steps of the clustering algorithm: (a) segmented image including identified small granular cells shown in red, (b) calculation of Euclidean distances to neighboring granules, (c) final result of the clustering illustrated in different colors.}
	\label{Cluster}
\end{figure*}	
According to \citet{2012ApJ...756L..27A}, mini-granules that form clusters and chains are situated in the intergranular lanes. To investigate this observational finding in the ANTARES simulation and to quantify the finding in a statistical analysis, we developed a clustering algorithm adapted to the segmented data.\\

As described in section \ref{sec:Automated granule detection in two dimensions}, at each time step a segmentation mask is produced. The clustering algorithm is based on these final segmentation masks. The masks are labeled and statistical information, such as the area and the centroid of each granules is derived. The clustering algorithm starts with the detection of small granules with diameters smaller than the threshold diameter of about 750~km\footnote{This diameter was determined by the global maximum in the probability density function of the diameters of granules (see section \ref{sec:Results}).}.\\

These small granules are labeled in ascending order and separated from the remaining larger granules. Each of them (shown in red in Fig.~\ref{Cluster}a) is examined separately. Euclidean distances between the centroids (shown in pink in Fig.~\ref{Cluster}b) of the investigated granular cells and all other granules are calculated. The four nearest granules (plotted in black in Fig.~\ref{Cluster}b) with a minimum distance to the examined one (red) are now analyzed (corresponding Euclidean distances are listed in Fig.~\ref{Cluster}b). Several criteria define the assignment of the investigated granular cells to a cluster:
\begin{itemize}
     \item if the examined granular cell is surrounded only by large cells and the nearest small cell is by definition located too far away (distance larger than 1.9 Mm), which is the case in Fig.~\ref{Cluster}b, the examined cell is classified as not belonging to a cluster (Fig.~\ref{Cluster}c), 
     \item if the examined granule has another small granule within its closest four neighbors, which does not already belong to a cluster, this neighboring small granule adopts the pixel value (label) of the examined granule, 
     \item if a neighboring small cell already belongs to a cluster, the examined cell receives the same label (pixel value) and is appended to the cluster. 
\end{itemize}
The algorithm results in an image consisting of labeled clusters, where each cluster itself consists again of several isolated small granules uniquely belonging to one of those clusters. Figure~\ref{Cluster}c indicates the clusters in different colors. Statistical properties, such as centroids and eccentricities of the clusters, are determined. The possibility to allocate clusters formed by several mini-granules enables us to quantify their spatial distribution in the field of view. The determination of the eccentricity\footnote{The eccentricity of a cluster is defined as the ratio of the distance between the foci of the ellipse enclosing the cluster and the length of the major axis. The value of the eccentricity is between 0 and 1.} of the clusters is used to determine the alignment of small granules within these clusters.

\section{Results}
\label{sec:Results}
\subsection{Physical characteristics of granules} 
The developed two-dimensional segmentation algorithm offers the possibility to retrieve statistical information of the segmented granular cells. For our analysis we used a two-hour data set consisting of 281 time steps with a temporal resolution of 30 seconds. Each time step contains an average of 93 granules, yielding a total statistical sample of about 20 500 granules. Physical properties, such as the vertical and horizontal velocity as well as the geometrical depth of the segmented granules are evaluated on surfaces of the optical depth $\tau = 1$. The emerging intensity of the segmented granules is analyzed at the top of the computational box. This can be understood as the intensity that reaches a virtual observer's telescope. At each time instant the segmentation mask is applied to the emerging intensity and to the profiles of the vertical and the horizontal velocities, to obtain characteristic parameters for granules, such as mean intensity, geometrical depth, and horizontal and vertical velocity values. These physical properties of identified granules are then analyzed with respect to their equivalent diameters via scatter plots.\\ 
\begin{figure*}
	\centering
		\includegraphics[width=1\textwidth]{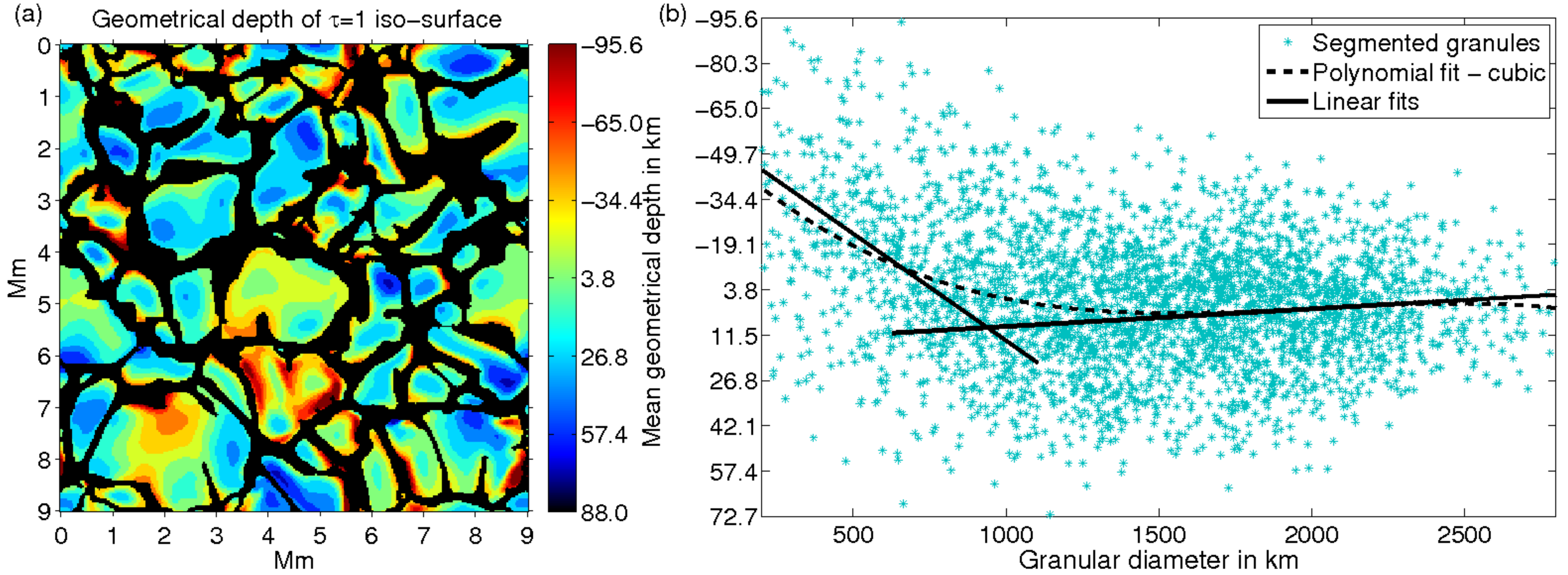}
		\caption{(a) Application of the segmentation mask to the geometrical depth of the iso-surface of $\tau = 1$ for a single frame. Red indicates the location below the calculated iso-surface of $\tau = 1$ in the convection zone and blue the location above this calculated level. (b) Granular diameter vs. the mean geometrical depth where $\tau = 1$ for the entire time series. The trend of the scatter is illustrated by linear fits (solid black lines) and a third-order polynomial fit (dashed dark-gray line).}
	\label{TauPos}
\end{figure*}

Figure~\ref{TauPos}a shows the segmentation mask colored according to the vertical position of the surface of optical depth unity. We can see that small granules are on average located deeper than the larger ones, which is also apparent in the scatter plot in Fig.~\ref{TauPos}b. The ordinate in Fig.~\ref{TauPos}b indicates the mean geometrical depth of the detected granules in the simulation box in km. Negative values indicate the location of a granule within the convective zone. The scatter is distributed in two distinctive regions with different trends. Thus, the applied two linear fits separate the scatter for the two distinctive regimes. The two linear fits intersect at a diameter of $\sim{1000}$~km. The group of small granules are apparently located at larger depths than regular granules, showing a trend of decreasing depth with increasing diameter, while larger granules are located at a more or less constant depth (only slightly increasing with size).\\

\begin{figure*}
	\centering
		\includegraphics[width=1\textwidth]{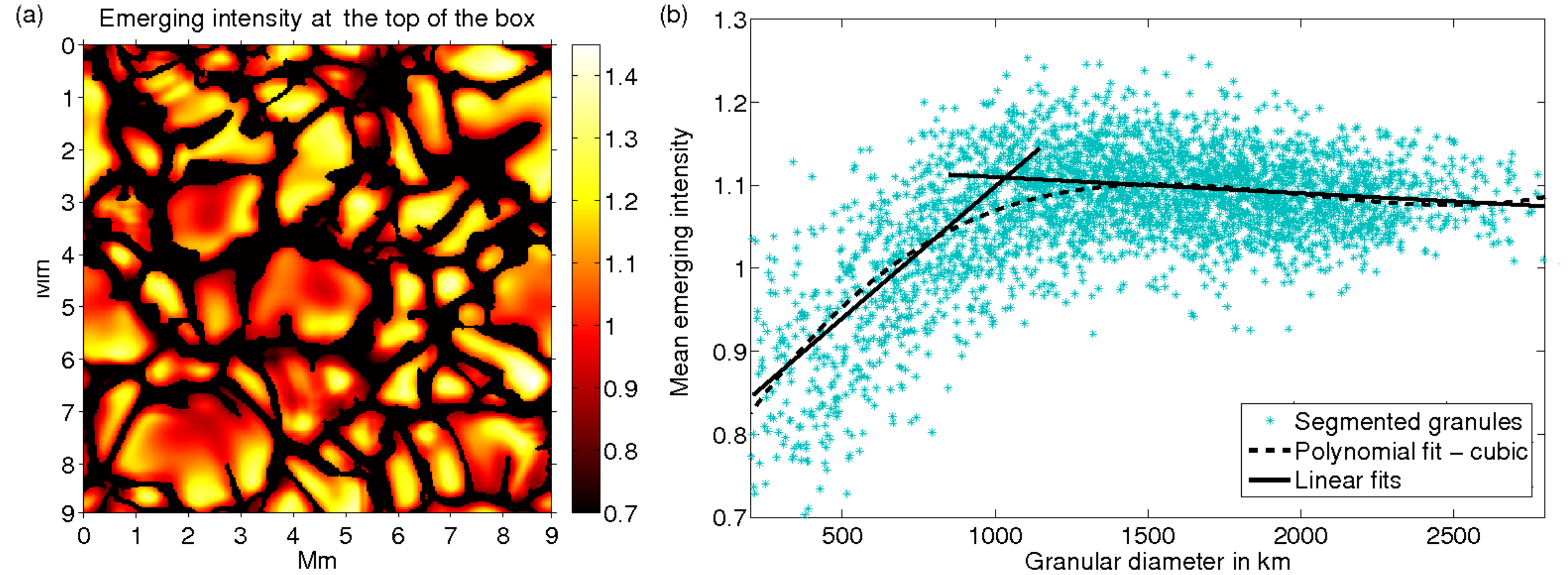}
		\caption{(a) Application of the segmentation mask to the normalized emerging intensity for a single frame. (b) Granular diameter vs. normalized mean emerging intensity of granules for the entire time series. The trend of the scatter is illustrated by linear fits (solid black lines) and a polynomial fit (dashed dark-gray line).}
	\label{Intensity}
\end{figure*}
\begin{figure*}
	\centering
		\includegraphics[width=1\textwidth]{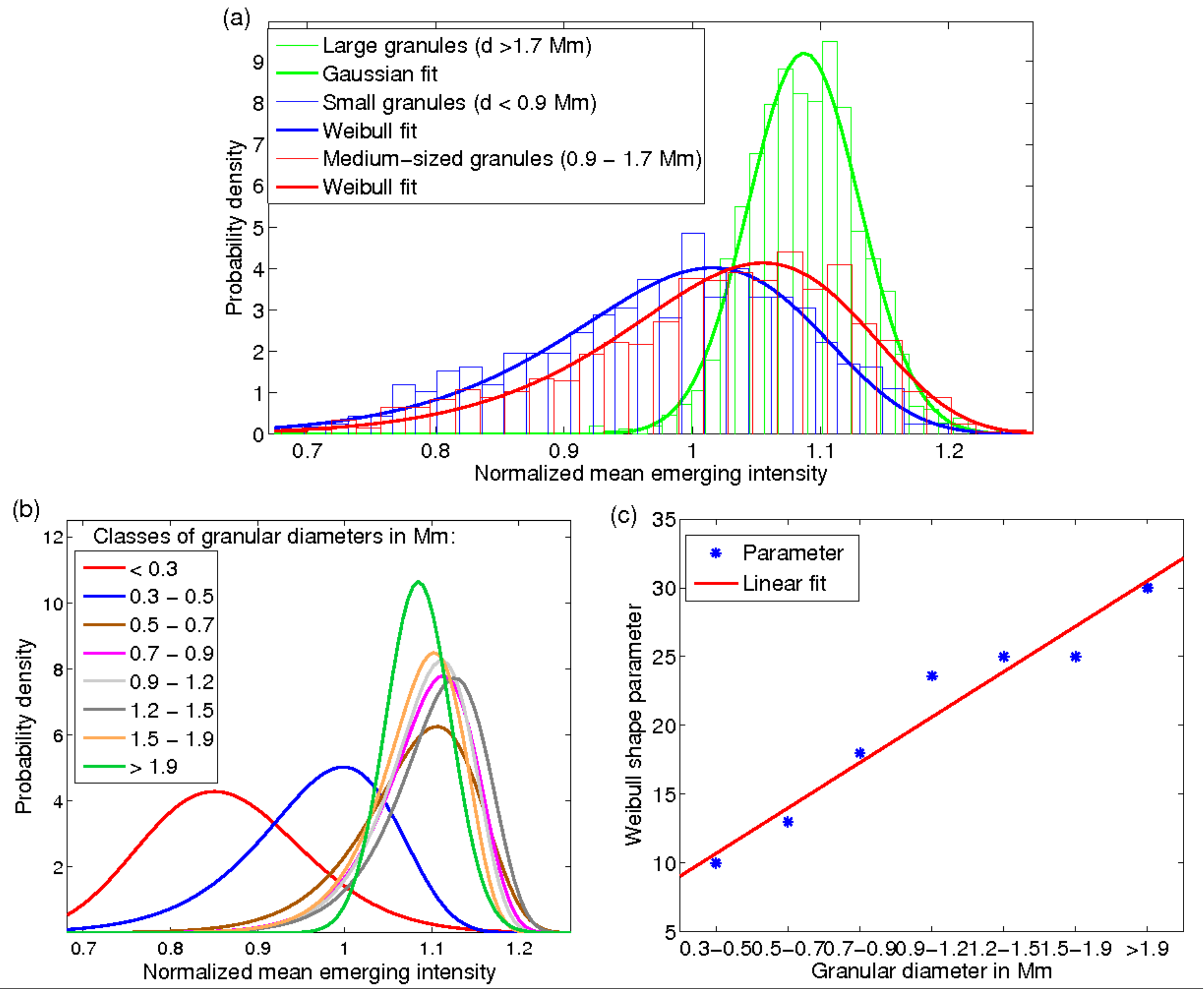}
		\caption{Analysis of normalized mean emerging intensity: (a) comparison of distributions of the normalized mean emerging intensity of large granules shown in green, medium-sized granules in red, and small granules in blue as well as their Gauss and Weibull fits, respectively, (b) family of Weibull curves derived from regrouping granules according to their diameters in classes, (c) shape parameter of the Weibull distributions derived from the intensity distribution family in (b).}
	\label{IntensityDists}
\end{figure*}
Figure \ref{Intensity}a shows the intensity distribution within granules and Fig.~\ref{Intensity}b the mean granular intensity versus its diameter. The intensity increases with increasing diameter up to a diameter of $\sim{1000}$~km, which is also verified by the slope of the third-order polynomial fit and the linear fits (dashed dark-gray and solid black lines, respectively). These fits also indicate a weakly decreasing trend for the intensities of granules larger than 1000~km in diameter. Interestingly, the split of granular features into two populations appears again, illustrated by two opposite linear trends, obtained for small and large granules.\\

In Fig.~\ref{IntensityDists}a the normalized emerging intensity distribution of small-, medium- and large-sized granules is analyzed separately. Large granules show regular Gaussian distributions while for small- and medium-sized granules the emerging intensity distributions differ from Gaussians but fit a Weibull distribution, as illustrated by the solid red and blue lines in Fig.~\ref{IntensityDists}a. As a consequence, we regrouped granules according to their diameters in several subpopulations and analyzed their intensity distributions separately. In this way, we obtained a family of Weibull curves with different shape parameters (see Fig.~\ref{IntensityDists}b). We plot these parameters versus the group diameters in Fig.~\ref{IntensityDists}c. The figure demonstrates that the best-fitting Weibull distribution shows a linear parameter dependence on the mean diameter of granules. The theoretical interpretation of the shape parameter, its dependence on granular diameters and the meaning of the Weibull distribution for the physics of the solar granulation has to be investigated in more detail and will be a possible topic for future study.\\
\begin{figure*}
	\centering
		\includegraphics[width=1\textwidth]{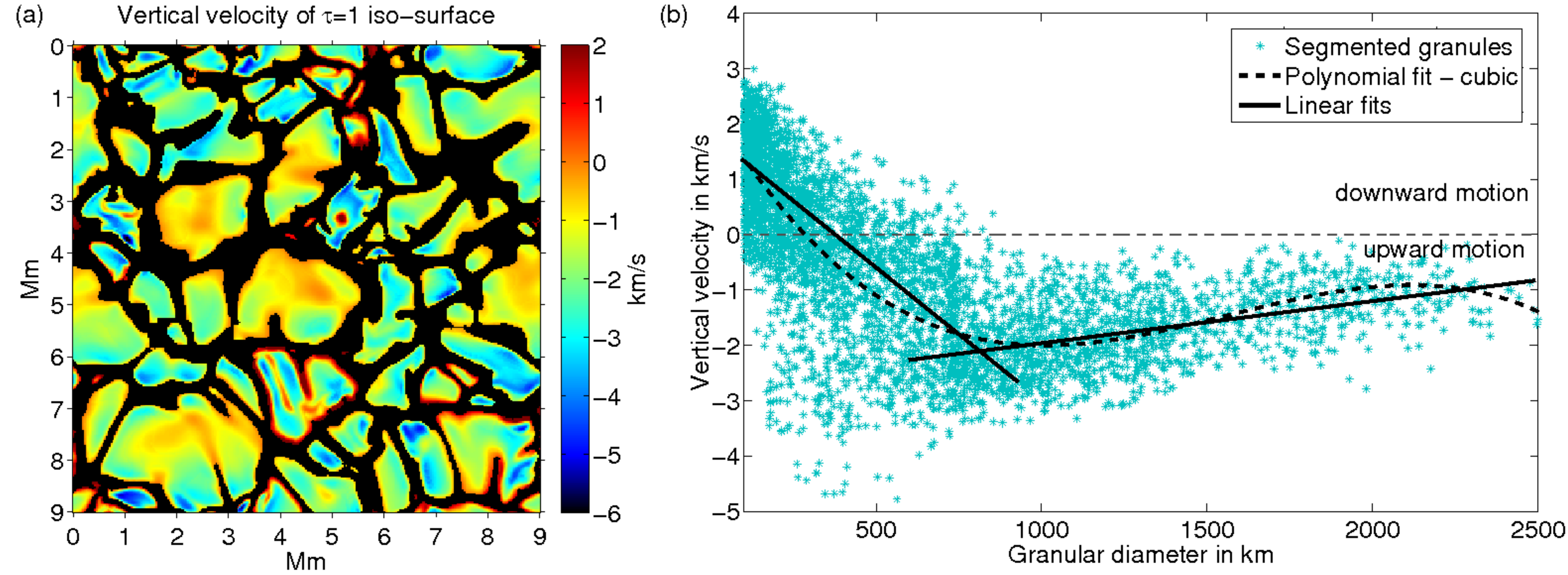}
		\caption{(a) Application of the segmentation mask to the vertical velocity on the $\tau_1$-iso-surface for a single frame. (b) Scatter plot of the mean vertical velocity of segmented granules vs. the granular diameter for the entire time series. The trend of the scatter is illustrated by two linear fits (solid black lines) and a third-order polynomial fit (dashed dark-gray line). Negative values correspond to upflowing plasma. }
	\label{VerticalVelocity}
\end{figure*}

We also analyzed the mean vertical velocity of segmented granules (Fig.~\ref{VerticalVelocity}a). The distribution of the mean vertical velocity of identified small granules exhibit a large scatter, which may be related to different stages of their evolution because they are predominantly found in deeper layers (see Fig.~\ref{VerticalVelocity}b). The polynomial and linear fits show a strong decrease of vertical velocity (upward moving plasma) with decreasing granular diameters and thus support the concept of the existence of two distinct populations.\\

\begin{figure*}
	\centering
		\includegraphics[width=1\textwidth]{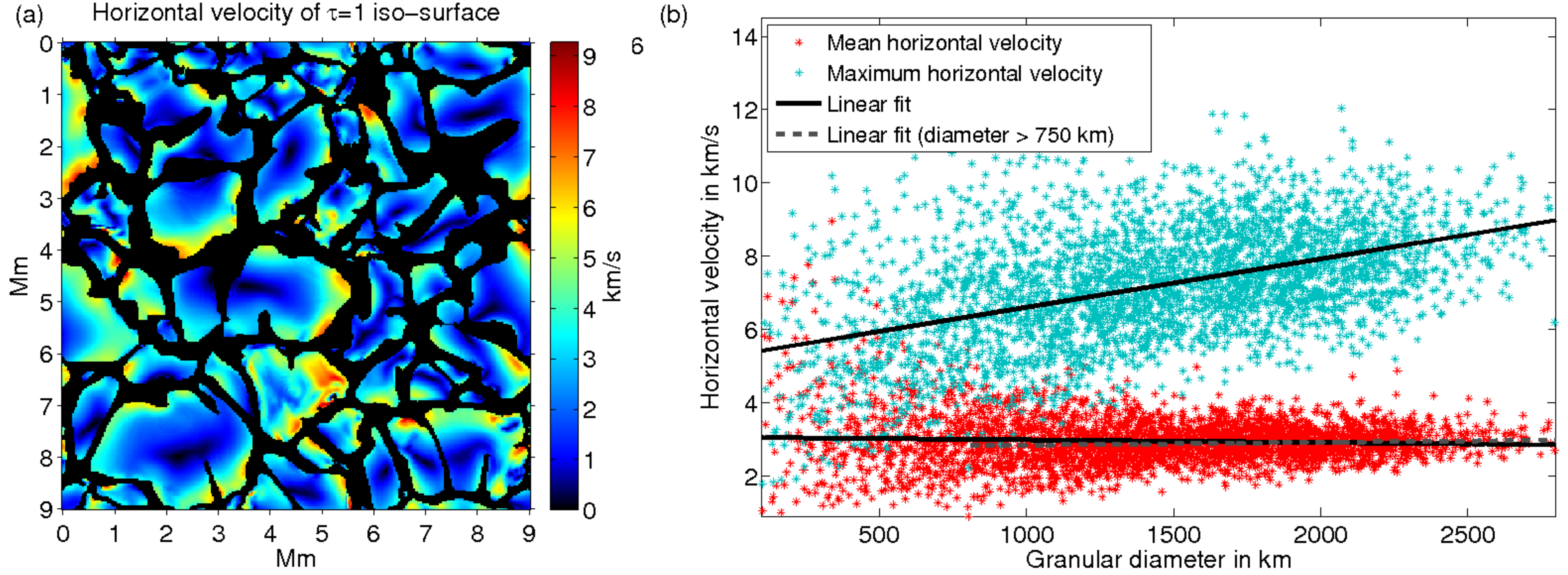}
		\caption{(a) Application of the segmentation mask to the horizontal velocity on the $\tau_1$-iso-surface for a single frame. (b) Scatter plot of granular diameter vs. mean horizontal velocity of segmented granules for the entire time series, shown in red and the maximum horizontal velocity in blue. The trends for granules across the whole range of diameters are illustrated by linear fits as solid black lines and for granules with diameters larger than 750 km by a linear fit as a dashed dark-gray line (only shown for the mean horizontal velocity).}
	\label{HorVelDist}
\end{figure*}

Applying the segmentation masks on the horizontal velocity profiles of the RHD model (see Fig.~\ref{HorVelDist}a) enables us to estimate the mean horizontal velocities (Fig.~\ref{HorVelDist}b, red), which reveal a slightly decreasing trend over the whole range of granular diameters (solid black line in Fig.~\ref{HorVelDist}b). We then excluded granules with diameters smaller than 750 km, because they exhibit a large scatter, and found for the remaining majority of granules a practically constant behavior, illustrated by the dashed approximately horizontal dark-gray fit. The maximum horizontal velocity (Fig.~\ref{HorVelDist}b, blue) is dependent on the granular size. As the diameter increases, the maximum horizontal velocity increases. This result coincides with findings in simulations \citep[e.g.][]{1989A&A...213..371S}. The increase in horizontal velocity to balance the larger plasma volume that ascends to the surface can be understood as a consequence of mass conservation \citep[see e.g.][]{2009LRSP....6....2N}. 

\subsection{Structural properties of granules}
\begin{figure*}
	\centering
		\includegraphics[width=1\textwidth]{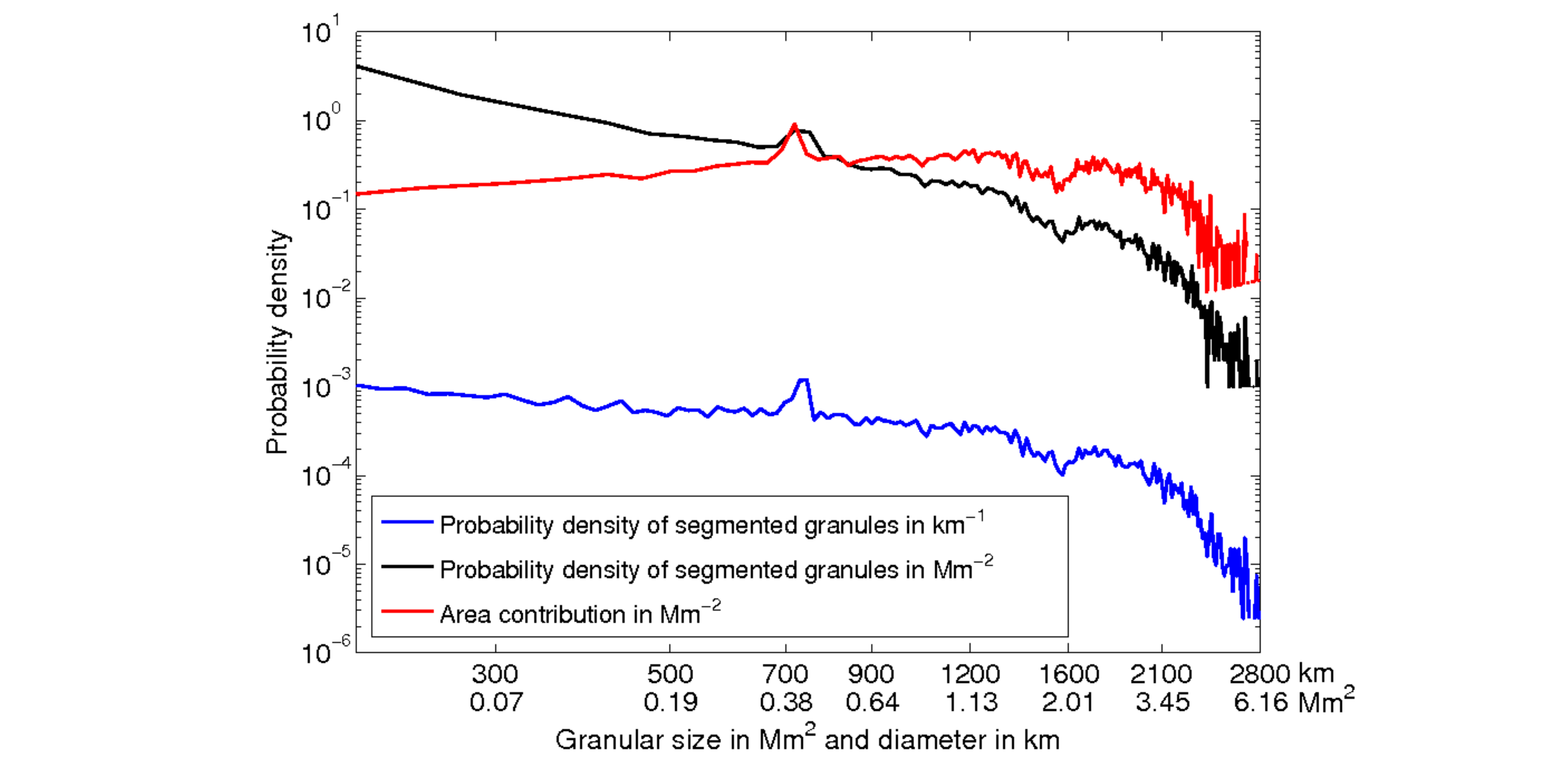}
		\caption{Area contribution function shown in red. The black curve indicates the probability density function of the granular area and the blue solid curv the equivalent diameter of segmented granules.}
	\label{AreaProb}
\end{figure*}
Figure~\ref{AreaProb} shows the area contribution function defined as the contribution of a given size of a granule to the total area of granules (\citet{1986SoPh..107...11R}). The function reveals a local minimum located in an area of $\sim{1.9}$~Mm$^{2}$, which corresponds to an equivalent diameter of $\sim{1500}$~km, and a global maximum in an area of $\sim{0.43}$~Mm$^{2}$ corresponding to an equivalent diameter of $\sim{750}$~km.\\

The probability density function of the equivalent diameter (blue line in Fig.~\ref{AreaProb}) shows an increase in the number of detected granules towards smaller scales. We found a less distinct change in the slope than \citet{2012ApJ...756L..27A}.\\

For our statistical analysis the distinct global maximum at a diameter of 750~km (Fig.~\ref{AreaProb}) was defined as the threshold value to discern between the population of small granules and larger ones. Interestingly, this value agrees well with the points of intersection of the linear fits applied to the vertical velocity and hence might have a real physical meaning.

\subsection{Cluster detection and analysis of small granules}
\begin{figure*}
	\centering
		\includegraphics[width=1\textwidth]{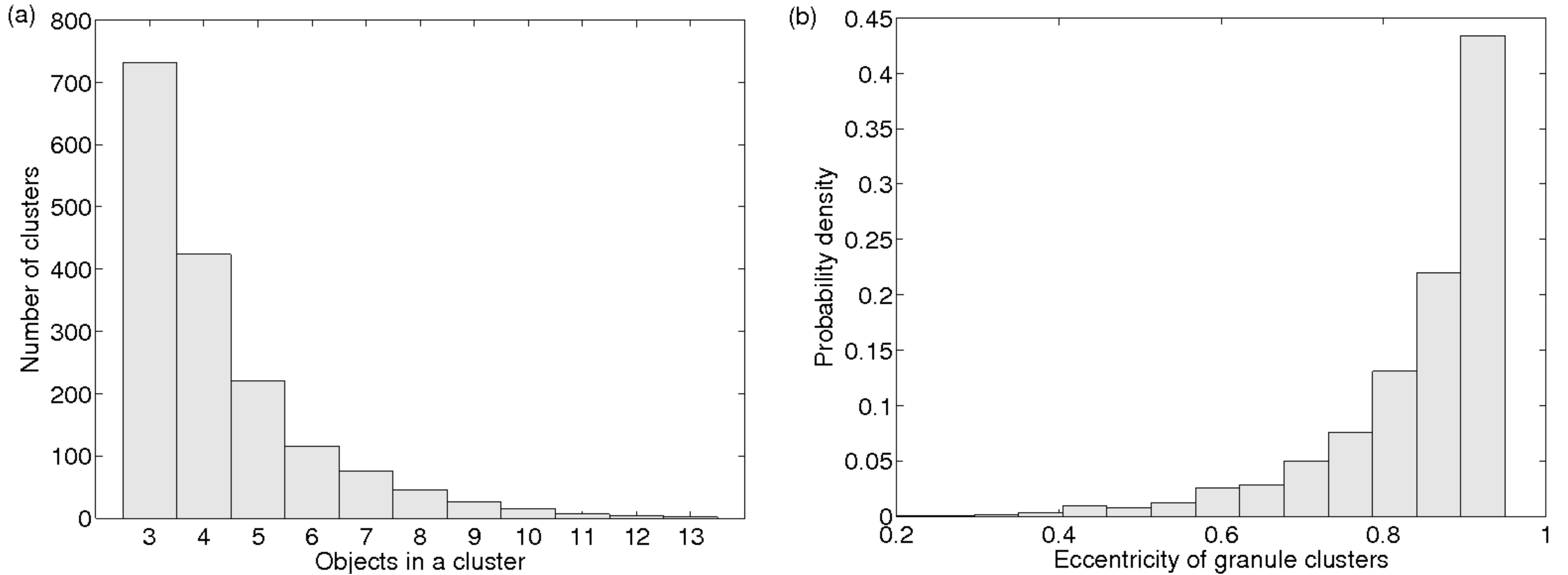}
		\caption{Analysis of the structural alignment of clusters: (a) histogram of the number of granules forming a cluster. (b) Probability density of the eccentricity of detected clusters that consist of more than two granules.}
	\label{ClustersAnalysis}
\end{figure*}

According to \citet{2012ApJ...756L..27A}, small granules often form clusters and chains. Our clustering algorithm enables us to quantify the spatial arrangement of these small granules with diameters smaller than 750~km. Figure~\ref{ClustersAnalysis} shows the distribution of the number of small granules in the clusters. We see that the majority of clusters consist of two granules. Because apparently two granules can only form lines and chains, we adjusted the distribution by neglecting all clusters formed by two granules and only considered clusters consisting of three or more granules. The resulting distribution of the eccentricities of clusters is displayed in Fig.~\ref{ClustersAnalysis}b. We found that small granules in clusters predominantly form chains. The main part of clusters feature an eccentricity higher than 0.9, that is close to 1 (alignment along a line; a value towards 0 would represent a circular alignment).

\section{Discussion and conclusions}
\label{sec:Discussion}
New telescopes and 3D simulations produce huge amounts of data, which make it necessary to develop algorithms for automated analysis. High-resolution simulations such as those obtained with the RHD simulation code ANTARES provide synthetic intensity maps and velocity profiles that can be compared with observations. Additionally, the simulation data allowed us to infer the characteristic physical quantities on any layers in the simulation box, which are not accessible to observations. The segmentation algorithm developed and described here, makes use of this additional information by combining a multiple-level thresholding routine applied to the vertical velocities with image-processing techniques, that use the emerging intensity profiles at the top of the computational box. 

\paragraph{Granular characteristics: }
The analysis of the horizontal and vertical velocity, and also all intensity distributions of the segmented granular cells, yealded different results for small granules than for granules larger than 1000~km. Thus we can apparently speak of two distinct populations of granules. The most distinct characteristics of small granules are their lower intensities and maximum horizontal velocities. The fits applied to the scatter plots of the emerging intensity as a function of granular size indicate a positive slope for granules with diameters smaller than $\sim{1000}$~km. On the other hand, a slight decrease is visible for the population of regular granules. A detailed study of the intensity distribution leads us to the conclusion that a Weibull distribution fits the results obtained for small and medium-sized granules very well, while the intensity distribution for larger granules (diameters larger than 1.9 Mm) can equally be approximated by a Gaussian. From the family of Weibull intensity distributions for classes of granules with different diameters we deduced a linear relationship between the Weibull shape parameter and mean granular class diameters. The larger the diameter, the more Gaussian-like the intensity distribution. These results agree in so far with findings in \citet{2011ApJ...743...58Y}, as they found that small granules do not satisfy Gaussian distributions, whereas the continuum intensity distribution of large granules is similar to a Gaussian.\\

The analysis of the vertical velocity as depicted in Fig.~\ref{VerticalVelocity}b illustrates a considerable scatter for small granules. Especially the smallest among them show downflowing motions, while larger granules are in a predominately upflowing state. This was confirmed by observations \citep[]{2011ApJ...743...58Y} and by results from simulations \citep[]{2000A&AS..146..267G}. To study the origin of these motions, detailed analysis of the temporal evolution of granules has to be carried out. 

\paragraph{Structural appearance and clustering: }
The area contribution function (Fig.~\ref{AreaProb}) agrees with findings in previous studies. We determined the global maximum at a diameter of $750$~km, which is smaller than the dominant scale. This  might be because of the high contrast of the data and the possibility to segment granules without loosing information due to the detectability of the smallest variations of intensity levels within detected granules. Another finding is that granules with diameters smaller than $\sim{750}$~km contribute significantly to the total area of granules.\\

The determined probability density function of the diameters of identified granules (Fig.~\ref{AreaProb}) differs from the results in \citet{2012ApJ...756L..27A}, where the probability density function shows a decrease from small diameters to diameters of up to 600~km. This probably results from over-segmentation because they used single thresholds to detect granules. While the authors stated that they excluded magnetic bright points\footnote{Small-scale magnetic flux concentrations that appear bright in filtergram observations at scales of $\sim{200}$~km in diameter.}, some of them might still have escaped the exclusion. Hence, these small granules may have contributed to their analysis, which might be another reason for the occurrence of many granules identified with a diameter smaller than 300~km, which cause the steep increase of their curve toward smaller sizes.\\

The eccentricity distribution derived for the clusters indicates that small granules form clusters that show chain-like alignments. This coincides well with observational results in \citet{2012ApJ...756L..27A}.

\paragraph{General remarks and outlook: }
When studying time series with an interval of 30 seconds as used in this analysis, small granules appear for only a few time steps and rarely reach the same intensity as the neighboring larger granules. This behavior suggests that small granules, which are situated in the intergranular lanes, are not evolving to the same geometrical height as larger granular cells. Movies suggest that small granules may not result from fragmentation of larger granular cells but instead evolve and dissolve in regions of intergranular lanes, rarely merging with other granules. For a detailed automated analysis of their evolution a higher time-cadence is necessary. An increase of the temporal resolution by an order of magnitude is desirable for the continuous investigation of the granular evolution. This will consequently lead to a better understanding of the evolution of the plasma upflow, manifested in two dimensions at photospheric levels, as granules.
In the future, a more complete understanding of the behavior of small granules will be gained by studying the 3D evolution of the upstreaming hot plasma plumes in the convection zone. For this purpose, we plan to develop a generalization of the segmentation algorithm that can be applied to three dimensions.

\begin{acknowledgements}
The research work was funded by the Austrian Science Fund (FWF): P23818 and P20762. D.U. is particularly greatful for the
special support given by project J3176 (Spectroscopical and Statistical Investigations on MBPs). The model calculations have been carried out at VSC (project P70068, H. Muthsam). A.H., B.L. and D.U. thank the {\"O}AD and ICD for financing a scientific stay at the Indian Institute of Astrophysics. R.K. thanks the ICD and {\"O}AD for financing a short research stay at the IGAM, Institute of Physics, of the University of Graz. The authors thank the anonymous referee for constructive comments that helped to improve the segmentation algorithm.
\end{acknowledgements}
\bibliographystyle{aa}
\bibliography{literature}

\begin{thebibliography}{28}
\expandafter\ifx\csname natexlab\endcsname\relax\def\natexlab#1{#1}\fi

\bibitem[{{Abramenko} {et~al.}(2012){Abramenko}, {Yurchyshyn}, {Goode},
  {Kitiashvili}, \& {Kosovichev}}]{2012ApJ...756L..27A}
{Abramenko}, V.~I., {Yurchyshyn}, V.~B., {Goode}, P.~R., {Kitiashvili}, I.~N.,
  \& {Kosovichev}, A.~G. 2012, \apjl, 756, L27

\bibitem[{{Bovelet} \& {Wiehr}(2001)}]{2001SoPh..201...13B}
{Bovelet}, B. \& {Wiehr}, E. 2001, \solphys, 201, 13

\bibitem[{{Collados}(2008)}]{2008SPIE.7012E..17C}
{Collados}, M. 2008, in Society of Photo-Optical Instrumentation Engineers
  (SPIE) Conference Series, Vol. 7012, Society of Photo-Optical Instrumentation
  Engineers (SPIE) Conference Series

\bibitem[{{Danilovic} {et~al.}(2008){Danilovic}, {Gandorfer}, {Lagg},
  {Sch{\"u}ssler}, {Solanki}, {V{\"o}gler}, {Katsukawa}, \&
  {Tsuneta}}]{2008A&A...484L..17D}
{Danilovic}, S., {Gandorfer}, A., {Lagg}, A., {et~al.} 2008, \aap, 484, L17

\bibitem[{{de Wijn} {et~al.}(2009){de Wijn}, {Stenflo}, {Solanki}, \&
  {Tsuneta}}]{2009SSRv..144..275D}
{de Wijn}, A.~G., {Stenflo}, J.~O., {Solanki}, S.~K., \& {Tsuneta}, S. 2009,
  \ssr, 144, 275

\bibitem[{{Florio} \& {Berrilli}(1998)}]{1998MmSAI..69..655F}
{Florio}, A. \& {Berrilli}, F. 1998, \memsai, 69, 655

\bibitem[{{Gadun} {et~al.}(2000){Gadun}, {Hanslmeier}, {Pikalov}, {Ploner},
  {Puschmann}, \& {Solanki}}]{2000A&AS..146..267G}
{Gadun}, A.~S., {Hanslmeier}, A., {Pikalov}, K.~N., {et~al.} 2000, \aaps, 146,
  267

\bibitem[{{Gonzalez} {et~al.}(2009){Gonzalez}, {Woods}, \&
  {Masters}}]{2009JBO....14b9901G}
{Gonzalez}, R.~C., {Woods}, R.~E., \& {Masters}, B.~R. 2009, Journal of
  Biomedical Optics, 14, 029901

\bibitem[{{Goode} {et~al.}(2010{\natexlab{a}}){Goode}, {Coulter}, {Gorceix},
  {Yurchyshyn}, \& {Cao}}]{2010AN....331..620G}
{Goode}, P.~R., {Coulter}, R., {Gorceix}, N., {Yurchyshyn}, V., \& {Cao}, W.
  2010{\natexlab{a}}, Astronomische Nachrichten, 331, 620

\bibitem[{{Goode} {et~al.}(2010{\natexlab{b}}){Goode}, {Yurchyshyn}, {Cao},
  {Abramenko}, {Andic}, {Ahn}, \& {Chae}}]{2010ApJ...714L..31G}
{Goode}, P.~R., {Yurchyshyn}, V., {Cao}, W., {et~al.} 2010{\natexlab{b}},
  \apjl, 714, L31

\bibitem[{{Grimm-Strele} {et~al.}(2013){Grimm-Strele}, {Kupka},
  {L{\"o}w-Baselli}, {Mundprecht}, {Zaussinger}, \&
  {Schiansky}}]{2013arXiv1305.0743G}
{Grimm-Strele}, H., {Kupka}, F., {L{\"o}w-Baselli}, B., {et~al.} 2013, ArXiv
  e-prints

\bibitem[{{Keil} {et~al.}(2010){Keil}, {Rimmele}, {Wagner}, \& {ATST
  Team}}]{2010AN....331..609K}
{Keil}, S.~L., {Rimmele}, T.~R., {Wagner}, J., \& {ATST Team}. 2010,
  Astronomische Nachrichten, 331, 609

\bibitem[{{Leitner} {et~al.}(2009){Leitner}, {Hanslmeier}, {Muthsam},
  {Veronig}, {L{\"o}w-Baselli}, \& {Obertscheider}}]{2009CEAB...33...69L}
{Leitner}, P., {Hanslmeier}, A., {Muthsam}, H.~J., {et~al.} 2009, Central
  European Astrophysical Bulletin, 33, 69

\bibitem[{{Lemmerer} {et~al.}(2010){Lemmerer}, {Hanslmeier}, {Muthsam}, \&
  {Leitner}}]{2010CEAB...34...39L}
{Lemmerer}, B., {Hanslmeier}, A., {Muthsam}, H., \& {Leitner}, P. 2010, Central
  European Astrophysical Bulletin, 34, 39

\bibitem[{{Lemmerer} {et~al.}(2012){Lemmerer}, {Utz}, {Hanslmeier},
  {K{\"u}hner}, {Grimm-Strele}, {Pauritsch}, {Thonhofer}, \&
  {Muthsam}}]{2012CEAB...36...29L}
{Lemmerer}, B., {Utz}, D., {Hanslmeier}, A., {et~al.} 2012, Central European
  Astrophysical Bulletin, 36, 29

\bibitem[{{Mundprecht} {et~al.}(2012){Mundprecht}, {Muthsam}, \&
  {Kupka}}]{2012arXiv1209.2952M}
{Mundprecht}, E., {Muthsam}, H.~J., \& {Kupka}, F. 2012, ArXiv e-prints

\bibitem[{{Muthsam} {et~al.}(2010){Muthsam}, {Kupka}, {L{\"o}w-Baselli},
  {Obertscheider}, {Langer}, \& {Lenz}}]{2010NewA...15..460M}
{Muthsam}, H.~J., {Kupka}, F., {L{\"o}w-Baselli}, B., {et~al.} 2010, \na, 15,
  460

\bibitem[{{Muthsam} {et~al.}(2007){Muthsam}, {L{\"o}w-Baselli},
  {Obertscheider}, {Langer}, {Lenz}, \& {Kupka}}]{2007MNRAS.380.1335M}
{Muthsam}, H.~J., {L{\"o}w-Baselli}, B., {Obertscheider}, C., {et~al.} 2007,
  \mnras, 380, 1335

\bibitem[{{Nordlund} {et~al.}(2009){Nordlund}, {Stein}, \&
  {Asplund}}]{2009LRSP....6....2N}
{Nordlund}, {\AA}., {Stein}, R.~F., \& {Asplund}, M. 2009, Living Reviews in
  Solar Physics, 6, 2

\bibitem[{{Roudier} \& {Muller}(1986)}]{1986SoPh..107...11R}
{Roudier}, T. \& {Muller}, R. 1986, \solphys, 107, 11

\bibitem[{{Schmidt} {et~al.}(2012){Schmidt}, {von der L{\"u}he}, {Volkmer},
  {Denker}, {Solanki}, {Balthasar}, {Bello Gonzalez}, {Berkefeld}, {Collados},
  {Fischer}, {Halbgewachs}, {Heidecke}, {Hofmann}, {Kneer}, {Lagg}, {Nicklas},
  {Popow}, {Puschmann}, {Schmidt}, {Sigwarth}, {Sobotka}, {Soltau}, {Staude},
  {Strassmeier}, \& {Waldmann }}]{2012AN....333..796S}
{Schmidt}, W., {von der L{\"u}he}, O., {Volkmer}, R., {et~al.} 2012,
  Astronomische Nachrichten, 333, 796

\bibitem[{{Schrijver} {et~al.}(1997){Schrijver}, {Hagenaar}, \&
  {Title}}]{1997ApJ...475..328S}
{Schrijver}, C.~J., {Hagenaar}, H.~J., \& {Title}, A.~M. 1997, \apj, 475, 328

\bibitem[{{Steffen} {et~al.}(1989){Steffen}, {Ludwig}, \&
  {Kruess}}]{1989A&A...213..371S}
{Steffen}, M., {Ludwig}, H.-G., \& {Kruess}, A. 1989, \aap, 213, 371

\bibitem[{{Steiner} {et~al.}(2010){Steiner}, {Franz}, {Bello Gonz{\'a}lez},
  {Nutto}, {Rezaei}, {Mart{\'{\i}}nez Pillet}, {Bonet Navarro}, {del Toro
  Iniesta}, {Domingo}, {Solanki}, {Kn{\"o}lker}, {Schmidt}, {Barthol}, \&
  {Gandorfer}}]{2010ApJ...723L.180S}
{Steiner}, O., {Franz}, M., {Bello Gonz{\'a}lez}, N., {et~al.} 2010, \apjl,
  723, L180

\bibitem[{{Utz} {et~al.}(2010){Utz}, {Hanslmeier}, {Muller}, {Veronig},
  {Ryb{\'a}k}, \& {Muthsam}}]{2010A&A...511A..39U}
{Utz}, D., {Hanslmeier}, A., {Muller}, R., {et~al.} 2010, \aap, 511, A39

\bibitem[{{Yu} {et~al.}(2011){Yu}, {Xie}, {Hu}, {Yang}, {Zhang}, \&
  {Wang}}]{2011ApJ...743...58Y}
{Yu}, D., {Xie}, Z., {Hu}, Q., {et~al.} 2011, \apj, 743, 58

\bibitem[{{Yurchyshyn} {et~al.}(2011){Yurchyshyn}, {Goode}, {Abramenko}, \&
  {Steiner}}]{2011ApJ...736L..35Y}
{Yurchyshyn}, V.~B., {Goode}, P.~R., {Abramenko}, V.~I., \& {Steiner}, O. 2011,
  \apjl, 736, L35

\bibitem[{{Zaussinger} \& {Spruit}(2013)}]{2013A&A...554A.119Z}
{Zaussinger}, F. \& {Spruit}, H.~C. 2013, \aap, 554, A119

\end{thebibliography}
\end{document}